\documentclass[aip,jcp,reprint,amsmath,amssymb]{revtex4-1}
\usepackage{graphicx}

\usepackage{color}

\newcommand*{\Fc}{F_{\text{c}}}
\newcommand*{\Fv}{F_{\text{v}}}

\newcommand*{\tc}{\tau_{\text{c}}}
\newcommand*{\Lc}{L_{\text{c}}}

\newcommand*{\dif}{\mathrm{d}}
\newcommand*{\od}[2]{\frac{\dif #1}{\dif #2}}

\begin{document}

\title[Capillary filling with giant wall slip]{Capillary filling with giant liquid/solid slip: dynamics of
  water uptake by carbon nanotubes}

\date{\today}

\author{Laurent Joly}\email{laurent.joly@univ-lyon1.fr}
\affiliation{LPMCN, Universit\'e de Lyon; UMR 5586 Universit\'e Lyon 1 et CNRS, F-69622 Villeurbanne, France}

\begin{abstract}
This article discusses the way the standard description of capillary
filling dynamics has to be modified to account for liquid/solid slip
in nanometric pores. It focuses in particular on the case of a large
slip length compared to the pore size. 
It is shown that the liquid viscosity does not play a role, and that the flow is
only controlled by the friction coefficient of the liquid at the 
wall. Moreover in the Washburn regime, the filling velocity does not
depend on the tube radius. Finally, molecular dynamics simulations
suggest that this standard description fails to describe the early stage of capillary filling of
carbon nanotubes by water, since viscous dissipation at the tube
entrance must be taken into account.
\end{abstract}

%\pacs{47.61.-k,47.55.nb,83.50.Rp,47.11.Mn} %showpacs

% 47.61.-k Micro- and nano- scale flow phenomena
% 47.55.nb Capillary and thermocapillary flows
% 83.50.Rp Wall slip and apparent slip
% 47.11.Mn Molecular dynamics methods 

% 47.15.G- Low-Reynolds-number (creeping) flows
% 47.56.+r Flows through porous media
% 68.08.-p Liquid-solid interfaces
% 83.50.Lh Slip boundary effects

%\keywords{Suggested keywords} %showkeys

\maketitle

\section{Introduction}
Dynamics of capillary rise is a standard case of coupling between
capillarity and %(low-Reynolds-number) 
hydrodynamics
\cite{dGBQ,Zhmud2000}. The filling behavior has been investigated
since the beginning of the 20th century
\cite{Lucas1918,Washburn1921,Bosanquet1923,Quere1997,Stange2003}. In
particular, the famous (Lucas-)Washburn law (neglecting the liquid
inertia) states that the filling velocity decreases as the inverse
square root of time.
But this old problem needs to be revisited as regards nanometric
pores \cite{Schoch2008,Sparreboom2009}. 
Even if continuum hydrodynamics
remains valid for simple liquids (including water) down to channel
sizes of a typical 1\,nm \cite{Bocquet2010}, surfaces will play an
increasing role. In particular,
deviations from the classical hypothesis of a no-slip boundary
condition (BC) at the liquid/solid interface have been predicted
theoretically and observed experimentally \cite{Bocquet2007}.
The simplest way to account for liquid/solid slip is the so-called 'partial slip' BC,
which links the slip velocity $v_{\mathrm{slip}}$ with the
shear rate at the solid surface $\partial_n v$: $v_{\mathrm{slip}} =
b\,\partial_n v$, in which the slip length $b$ is the depth inside the
solid where the linear extrapolation of the velocity profile vanishes
(Fig. \ref{fig:system}.a).
Let us emphasize here that the slip length, though it has a simple
geometric meaning, is not a fundamental property of the liquid/solid
interface.
Indeed, the partial slip BC physically stems from the continuity of tangential
stress at the wall: the viscous shear stress exerted
by the liquid on the wall $\eta\,\partial_n v$ ($\eta$ being the
shear viscosity of the liquid) is equal to the friction force suffered by the liquid
from the wall, which can be written as $F/{\cal A} = -
\lambda \,v_{\mathrm{slip}}$ where $\lambda$ is the liquid/solid
friction coefficient and ${\cal A}$ the contact area. 
The slip length is accordingly a combination of the \emph{bulk} liquid
viscosity and the \emph{interfacial} friction coefficient: $b=\eta/\lambda$. 
For simple liquids %(\textit{e.g.} water) 
on smooth surfaces, slip
lengths up to a few tens of nanometers have been experimentally 
measured \cite{Bocquet2007}. Liquid-solid slip is therefore
expected to significantly affect flows in channels of nanometric
size, even when continuum hydrodynamics remains valid. 

% previous works
In recent years, many works have investigated capillary filling at
the nanoscale, emphasizing the roles of dynamic contact angle, liquid
inertia, and liquid-solid slip \cite{Marmur1988, Martic2004,
  Supple2004, Dimitrov2007, Huber2007, Cupelli2008, Schebarchov2008,
  Gruener2009, Ahadian2010, Chen2010, Stukan2010, Wu2010,Phan2010,
  Schebarchov2011,Qin2011}. 
% In particular there has been some confusion on the correct way to
% account for liquid-solid slip \cite{Dimitrov2007, Schebarchov2008,
%  Chen2010, Schebarchov2011}, suggesting a need for clarification. 
%
In this context, recent experiments
\cite{Majumder2005,Holt2006,Whitby2008,Du2011} and numerical simulations
\cite{Thomas2008,Falk2010} have reported slip lengths of water (and
other liquids) in carbon nanotubes (CNT) much larger than the tube radius. 
This article seeks to investigate the way this quite special condition
modifies the dynamics of capillary filling. 
Firstly, the full equations describing capillary filling
with a partial slip boundary condition will be derived and solved. The
article will then focus on the case of CNTs where giant liquid/solid slip has been
reported. Finally this standard analytical model will be tested
against molecular dynamics simulations.

\section{Analytical model}

Before turning to the specific case of water filling a CNT, the
equation describing capillary filling with a 
partial slip boundary condition, taking into account the liquid
inertia, will be derived and solved. This model is formally equivalent
to the one introduced by Supple and Quirke \cite{Supple2004}, 
with the difference that the equation will here be derived in the
framework of continuum hydrodynamics, as we have discussed that
liquid/solid slip could be significant for channel sizes where
continuum hydrodynamics remained valid.

%
%%%%% figure: model %%%%%
\begin{figure}
\includegraphics[width=8.6cm]{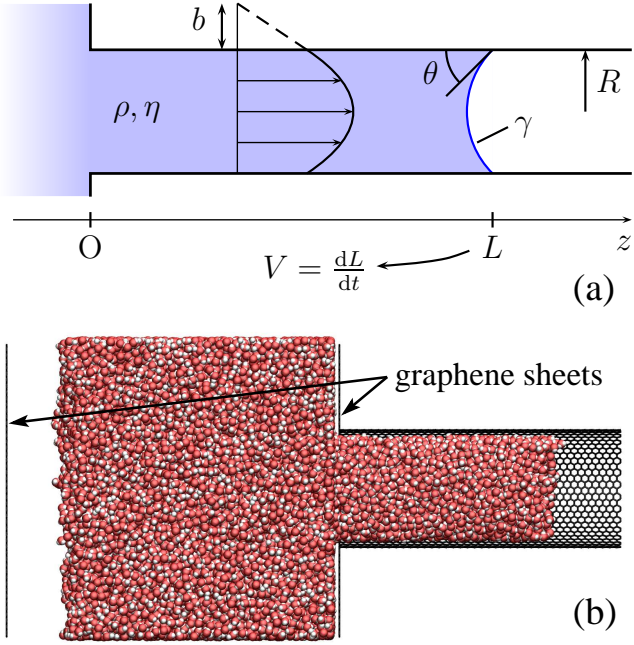}
\caption{(a) Schematics of the system considered. (b) Snapshot of a
  typical system used for the Molecular Dynamics simulations, created
  by using the VMD molecular visualization program \cite{VMD}.}
\label{fig:system}
\end{figure}
%%%%% figure: model %%%%%
%
%\emph{Equation of motion--} 
Let's consider a cylindrical pore of radius $R$, in
contact with a reservoir of liquid (density $\rho$, viscosity $\eta$),
as illustrated in Fig. \ref{fig:system}.a.
% (the slit pore case will be treated in appendix). 
The liquid surface tension is denoted $\gamma$, and its contact angle on the pore wall $\theta$ (it is assumed that the liquid is wetting the pore, \textit{i.e.} $\theta < \pi/2$).
The distance of the meniscus from the pore entrance is denoted $L$. Therefore the velocity of the meniscus $V$, equal to the average liquid velocity inside the pore, is related to $L$ through: $V = \od{L}{t}$.
The filling dynamics results from the competition between a capillary force at the contact line $\Fc$, driving the motion of the liquid inside the pore, and a viscous friction force at the wall $\Fv$, slowing the liquid motion (gravity can usually be neglected at small scales). One can write the equation of motion for the liquid inside the pore: 
\begin{equation}\label{eq:motion}
\od{(MV)}{t} = \Fc + \Fv , 
\end{equation}
where $M = \rho \pi R^2 L$ is the mass of liquid inside the pore.
We will denote $\Delta \gamma = \gamma_{\mathrm{SG}} -
\gamma_{\mathrm{SL}} = \gamma \cos \theta$ the capillary force per
unit length at the contact line, where $\gamma_{\mathrm{SG}}$ and
$\gamma_{\mathrm{SL}}$ are the solid-gas and solid-liquid surface tensions.
The total capillary force is then simply given by:
\begin{equation}\label{eq:Fc}
\Fc = 2\pi R \Delta \gamma .
\end{equation}
The friction force is given by the Poiseuille law \cite{dGBQ,Zhmud2000}, modified to take
into account liquid/solid slip \cite{Washburn1921,Schebarchov2008, Chen2010, Schebarchov2011}: 
%If the flow inside the pore remains laminar, the velocity profile must assume a parabolic shape: $v_z(r) = \Vm (1 - Ar^2)$, where $\Vm$ and $A$ are two arbitrary constants. Imposing a partial slip BC with a slip length $b$ at the wall: 
%\begin{equation}\label{eq:BC}
%v_z(r=R) = -b \left. \pd{v_z}{r} \right|_{r=R} , 
%\end{equation}
%one gets: 
%\begin{equation}\label{eq:vz}
%v_z(r) = \Vm \left( 1 - \frac{r^2}{R^2+2bR}  \right).
%\end{equation}
%The average velocity of the liquid inside the pore $V$ can then be related to the maximum velocity $\Vm$: 
%\begin{equation}\label{eq:V}
%V = \frac{1}{\pi R^2} \int_0^R v_z(r)\ 2\pi r \dif r = \frac{\Vm}{2} \frac{R^2+4bR}{R^2+2bR}.
%\end{equation}
%The friction force integrated over the entire liquid/solid interface is given by: 
%\begin{equation}
%\Fv = \int_0^L \dif z \int_0^{2\pi} R \dif \theta\ \eta \left. \pd{v_z}{r} \right|_{r=R} .
%\end{equation}
%Knowing the velocity profile \eqref{eq:vz}, and expressing $\Vm$ as a function of $V$ thanks to Eq. \eqref{eq:V}, this can be integrated to:
%\begin{equation}\label{eq:Fv}
%\Fv = -4\pi \eta L \Vm \frac{R^2}{R^2+2bR} = -\frac{8\pi \eta L V}{1+4b/R} .
%\end{equation}
\begin{equation}\label{eq:Fv}
\Fv = -\frac{8\pi \eta L V}{1+4b/R} .
\end{equation}
The equation of motion \eqref{eq:motion} can finally be re-written: 
\begin{equation}\label{eq:full}
\boxed{\od{(\rho \pi R^2 L V)}{t} = 2\pi R \Delta \gamma - \frac{8\pi \eta L V}{1+4b/R} }.
\end{equation}
This non-linear equation admits the following solution
\cite{Supple2004}:  
\begin{equation}\label{eq:fullsolution}
\frac{L (t)}{\Lc} = \left( \frac{t}{\tc} + \frac{e^{-2 t/\tc}-1}{2}
\right)^{1/2} , 
\end{equation}
with: 
\begin{equation}\label{eq:tc}
\tc = \frac{\rho R^2}{4\eta} \left( 1 + \frac{4b}{R} \right) , 
\end{equation}
and: 
\begin{equation}\label{eq:Lc}
\Lc = \frac{\left(2\gamma \rho R^3 \right)^{1/2}}{4\eta}  \left( 1+\frac{4b}{R} \right) . 
\end{equation}
One can easily check that in the 
long time limit, where the liquid inertia can be neglected compared to the
viscous friction inside the tube, Eq. \eqref{eq:fullsolution} simplifies into the
well-known Washburn law  \cite{Washburn1921,Schebarchov2008, Chen2010,
  Schebarchov2011}:
\begin{equation}\label{eq:washburn}
L^2(t) = \frac{\Delta \gamma R}{2\eta} \left( 1+\frac{4b}{R} \right) t .
\end{equation}
% One can note that the correction factor is the same one
% giving the amplification of flow rate in a cylindrical duct for a
% given pressure drop. This can be easily understood as in the viscous
% regime, capillary filling can be treated as a
% capillary-pressure-induced Poiseuille flow.
On the contrary, in the short time limit viscous friction can be
neglected compared to inertia. Eq. \eqref{eq:fullsolution} simplifies
into: 
\begin{equation}\label{eq:inertial}
L(t) = \left( \frac{2\Delta \gamma}{\rho R} \right)^{1/2} t. 
\end{equation}
Of course neither the viscosity $\eta$ nor the slip length $b$ appear
here since the friction term has been neglected.
One could finally note that even this inertial regime does not appear instantaneously, essentially
because the meniscus shape changes when the liquid enters into the
pore. But it can be shown that the meniscus adopts a stationary shape
for $L \sim R/2$ (from \cite{Stange2003}). Therefore, the pre-inertial
regime disappears extremely fast in nanometric pores.

% %
% %%%%% figure: CNT %%%%%
% \begin{figure}
% \includegraphics[width=8.6cm]{fig2}
% \caption{(a) Full line: Evolution of the unitless filling length
%   $L^\star = L/\Lc$ with unitless time $t^\star = t/\tc$ (full
%   solution); Dashed lines: Short-time/inertial and long-time/viscous
%   limits. 
% (b) Transition length $\Lc$ between inertial and viscous regimes, for
% water in a CNT, as a function of the tube radius $R$. Full line:
% analytical expression in the plug-flow approximation, for
% $\lambda=1.2\times 10^4$\,Ns/m$^3$ (from \cite{Falk2010}); Dashed line:
% no-slip limit. For very small tubes, the friction coefficient
% $\lambda$ decreases: Data from \cite{Falk2010} have been used. 
% % The 3 colors correspond to $R_{\text{hydro}} = R$,
% % $R-\frac{\sigma_{\text{OC}}}{2}$,and $R-\sigma_{\text{OC}}$. 
% Inset: Transition time $\tc$, with the same notations.}
% \label{fig:CNT}
% \end{figure}
% %%%%% figure: CNT %%%%%
% %

% water in CNT
\emph{Water inside CNT--} Recent experiments
\cite{Majumder2005,Holt2006,Whitby2008,Du2011} and numerical
simulations \cite{Thomas2008,Falk2010} have reported surprisingly
large slip lengths of water (and other liquids) in carbon nanotubes. The first experiments \cite{Majumder2005} indicated slip lengths up to tens of micrometers, even if more recent experimental \cite{Holt2006} and numerical \cite{Falk2010} works point to a smaller effect, with $b$ on the order of a few hundredth of nanometers. 
When the slip length is much larger than the pore size: $b \gg R$, the
flow inside the pore will be almost plug-like, namely with a constant
velocity profile. In this regime, there is almost no shear inside the
liquid, and the viscosity is not expected to play a role. In fact, the
viscous term $\Fv$ will only depend on the interfacial friction
coefficient $\lambda$. This can be seen if one take the limit $b/R \gg
1$ in Eq. \eqref{eq:Fv}:  
\begin{equation}
\Fv = -\frac{8\pi \eta LV}{4b/R} = -2\pi R \left(\frac{\eta}{b}\right) LV = -2\pi R \lambda LV.
\end{equation}
But this expression can be found directly under the assumption of a perfect plug flow, where the velocity of the liquid is everywhere equal to its average value $V$. The friction force over the liquid/solid contact area $\mathcal{A}=2\pi R L$ is then simply: 
\begin{equation}\label{eq:Fv_plug}
\Fv = -\mathcal{A} \lambda V = -2\pi R \lambda LV. 
\end{equation}
Generally the ratio $\eta/(1+4b/R)$ can be replaced by $\lambda R/4$
in the equation of motion \eqref{eq:motion}, its solutions
Eqs. \eqref{eq:washburn} and \eqref{eq:fullsolution}, and the
expression of the transition time \eqref{eq:tc} and transition length
\eqref{eq:Lc}.
In particular the Washburn law (in the viscous regime) becomes: 
\begin{equation}\label{eq:washburnplug}
\boxed{L^2(t) = \frac{2 \Delta \gamma}{\lambda} t}.
\end{equation}
In the plug-flow case, viscosity does not play a role,
and the flow is only controlled by the friction coefficient. Furthermore, the
filling dynamics does not depend on the tube radius anymore. This
important difference could be used as a clear experimental signature
of a plug flow in the viscous regime.

To compute orders of magnitude, one also needs to quantify the driving
force $\Fc$. This is far from trivial: Whereas measurements of water
contact angle on highly oriented pyrolytic graphite converge to a value of $86\,^\circ$
\cite{Alexiadis2008,Werder2003}, corresponding to $\theta =
96\,^\circ$ for water on a single graphene sheet (hence a non-wetting
situation, where water should not invade CNT), there are experimental
evidences that water fills at least the very small CNTs
\cite{Kolesnikov2004,Cambre2010,Qin2011,Kyakuno2011,Pascal2011}. At these scales however, it is hard to
apply macroscopic concepts of capillarity like surface tension and
contact angle \cite{Honschoten2010}. But if water indeed fills CNT,
$\Fc$ will be positive and one can estimate that it will be at most on the order of $2\pi R \gamma$.

% Following these considerations, one can estimate the order of
% magnitude for the transition time between inertial and viscous regime:
% $\tc = \rho R / \lambda$. Interestingly, $\tc$ scales now as $R$ in
% the plug-flow regime, whereas it scaled as $R^2$ in the no-slip
% regime. 
% The (maximum) corresponding filling length is given by: $\Lc \sim (2\rho R
% \gamma)^{1/2} / \lambda$. 
% For water ($\rho = 1\,\mathrm{g/cm^3}$, $\gamma = 0.072$\,N/m),
% Fig. \ref{fig:CNT}.b presents the evolution of $\Lc$ and $\tc$ with
% the tube radius, computed using the values of friction coefficient
% reported in \cite{Falk2010}. 
% %
% The computed transition time and length scales are too large to be
% reached with molecular dynamics simulations (especially with water,
% involving the computation of long-range Coulomb interactions), and yet
% quite small from the experimental point of view. One can conclude that
% molecular simulations should mainly have access to the inertial regime,
% whereas experiments will only be able to investigate the viscous
% regime. Therefore it will be very hard to confront both approaches
% on this kind of system. To end on an optimistic note, the extremely
% small friction coefficients measured numerically in (5,5) and (6,6)
% CNT \cite{Falk2010} could extend the inertial domain up to time and
% length scales accessible by experiments (Fig. \ref{fig:CNT}.b), even if these very
% small CNT would be quite delicate to manipulate.

\section{Molecular Dynamics simulations and discussion}

We considered a water reservoir in contact with an initially empty
CNT (Fig. \ref{fig:system}.b). The CNT length was 10\,nm, with radii
ranging between 0.514 and 1.87\,nm. The reservoir was bordered with two
graphene sheets, at a distance large enough to ensure that the liquid
water was always in equilibrium with its vapor. Periodic boundary
conditions in the two directions perpendicular to the tube axis ensured
that the water surface remained planar. Consequently, the pressure in
the water reservoir always remained constant, at a value extremely
close to zero.

The Amber96 force field \cite{AMBER96} was used, with TIP3P
water and water-carbon interaction modeled by a Lennard-Jones
potential between oxygen and carbon atoms, with parameters
$\varepsilon_{\mathrm{OC}} = 0.114$\,kcal/mol and $\sigma_{\mathrm{OC}} = 3.28$\,\AA.
The tabulated density, surface tension and viscosity of TIP3P water
at 300\,K and liquid/vapor coexistence are respectively: $\rho =
0.980$\,g/cm$^3$, $\gamma = 0.0523$\,N/m (from \cite{Vega2007}), and $\eta =
0.321$\,mPa\,s (from \cite{Gonzalez2010}). The contact angle of
water on a single graphene sheet was measured independently to be
$\theta = 57^\circ$ for this model.
The simulations were carried out using LAMMPS
\cite{LAMMPS}. Long-range Coulomb forces were computed using the
particle-particle particle-mesh (PPPM) 
method; A timestep of 2\,fs was used, unless specified.
The positions of the carbon atoms were fixed (simulations with
flexible and fixed walls were shown to give similar results for the
statics and friction of confined liquids
\cite{Alexiadis2008,Thomas2009,Werder2003}). 
Water molecules were kept at a constant temperature of 300\,K using a
Nos\'e-Hoover thermostat, applied only to the degrees of freedom
perpendicular to the tube axis, with a damping time of
200\,fs. Alternatively a Dissipative Particle Dynamics thermostat was
used (see details in the following).

\begin{figure}
\includegraphics[width=8.6cm]{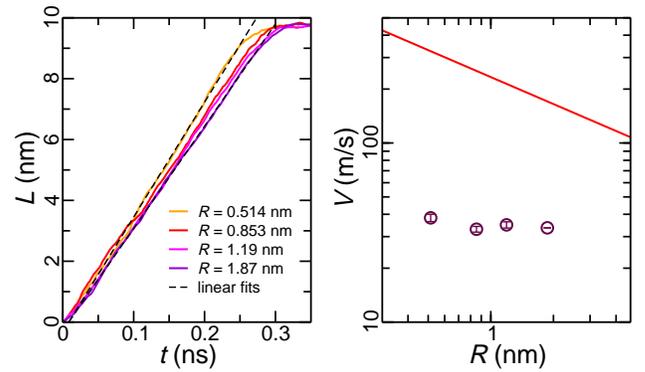}
\caption{(a) Evolution of the filling length $L$ with time $t$, for
  different tube radii. Dashed lines represent two examples of linear fit
  used to compute the filling velocity $V$. (b). Filling velocity $V$ as a function of
  tube radius $R$: Simulation data (circles) and prediction of the
  standard model (line), using Eq. \eqref{eq:inertial}.}
\label{fig:MD}
\end{figure}

Water molecules are initially disposed on a simple cubic lattice with
equilibrium density, so firstly the system is equilibrated during
typically 120\,ps, with a plug at the tube entrance to prevent water from
entering. Then the plug is removed and the evolution of the water mass inside
the tube is recorded as a function of time. To compute the filling
length, the mass is divided by the linear density of water inside the
tube, measured once the tube is completely filled. Results for
different tube radii are presented in Fig. \ref{fig:MD}.a.
%
%%%%% figure: MD %%%%%
%%%%% figure: MD %%%%%
%
It appears that $L$ scales linearly with time, as expected from the
model in the short-time limit. In Fig. \ref{fig:MD}.b, the filling velocities (slope of the
$L(t)$ curves) for different radii are compared to the theoretical
prediction of Eq. \eqref{eq:inertial}. The theoretical prediction
fails to describe the numerical data both quantitatively (simulated
velocities are much smaller than expected) and
qualitatively: While the model predicts that $V$ should decrease as
$R^{-1/2}$, the simulation shows that $V$ does not
depend significantly on the tube radius.

\begin{figure}
\includegraphics[width=8.6cm]{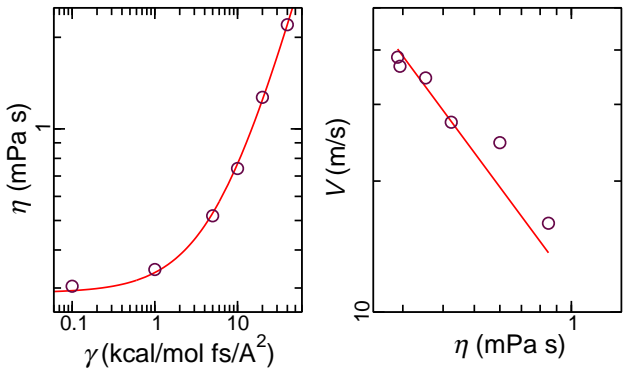}
\caption{(a) Calibration of viscosity $\eta$ versus DPD coefficient
  $\gamma$: MD results (circles) and fitting curve (line), see text
  for detail. (b) Filling velocity $V$ as a function of liquid
  viscosity $\eta$: MD results (circles) and fit with an inverse
  function (line).}
\label{fig:viscosity}
\end{figure}

While the quantitative discrepancy could be explained by a reduced 
surface tension (due to curvature and interaction with
the surface), the invariance of $V$ with the radius $R$ seems to
point toward another effect, for instance a drag force missing in the model. 
Even if viscosity is not expected to play a role inside the tube due to the
very large slip length, its influence was nevertheless tested, using a Dissipative Particle
Dynamics (DPD) thermostat \cite{Groot1997}. This amounts to adding pairwise
interactions between atoms, with a dissipative force depending on the
relative velocity between each pair and a random force with a Gaussian
statistics. 
% The dissipative term in DPD depends on the relative velocity between
% atoms, thus
This method has the advantage of preserving hydrodynamics.
Furthermore, the amplitude of the dissipative term can be tuned
 to modify the liquid viscosity without changing its static
properties.
The viscosity dependency on the dissipative term was carefully
calibrated in independent simulations of shear
flow of water between two silica surfaces (Fig. \ref{fig:viscosity}.a).
For the largest dissipative terms, the timestep had to be reduced down to 0.5\,fs.
%
%%%%% figure: viscosity %%%%%
%%%%% figure: viscosity %%%%%
%

The filling velocity was then measured for a given CNT, with radius 1.87\,nm,
varying the viscosity of the DPD-thermostated-water. On
Fig. \ref{fig:viscosity}.b, one can observe a clear influence of
viscosity on the filling velocity.
Due to the very large slip length of water inside CNT, friction at the wall
should be completely negligible for the filling lengths
simulated. Moreover, in such a plug-flow regime, the liquid moves as a 
block everywhere inside the tube. Therefore, there should be no dissipation near 
the contact line, and consequently no dynamical change of the contact angle.
Yet, even in the absence of friction inside the tube, viscous
dissipation can occur at the tube entrance 
due to the contraction of the streamlines
\cite{Sampson1891,Johansen1930,Weissberg1962,Nierop2008,Nierop2009,Thomas2009,Sisan2011}. 
The power dissipated by viscosity can be estimated as: $P_{\mathrm{v}}
= \frac{\eta}{2} \int_{\mathcal{V}} \mathrm{d}\mathcal{V} (\partial_i
v_j + \partial_j v_i)^2$. The only length scale related to the
entrance of the liquid inside the tube is the tube radius, hence:
$P_{\mathrm{v}} \propto \eta R^3 (V/R)^2 \propto \eta R V^2$. Neglecting the
kinetic energy of the liquid (inertial term): $P_{\mathrm{k}} \propto
\rho R^2 V^3$, the viscous dissipation must be equal to the power of
the driving capillary force: $P_{\mathrm{c}} = 2\pi R \Delta \gamma
V$. Finally $P_{\mathrm{c}} = P_{\mathrm{v}} \Rightarrow V \propto
\Delta \gamma / \eta$. Therefore, if the filling is controlled by
viscous dissipation at the entrance, the velocity should not depend on
the tube radius (as observed in Fig. \ref{fig:MD}.b), and 
scale with the inverse of viscosity. One can check in
Fig. \ref{fig:viscosity} that the numerical results for $V$ versus
$\eta$ are correctly fitted with an inverse function.
To conclude, MD simulations show that, for water filling CNTs, inertia
is negligible with regard to viscous dissipation at the entrance.
The standard description of capillary filling presented
earlier must thus be modified.

In this paragraph, a simple analytical description of the
full capillary filling dynamics, taking into account entrance effects
(but neglecting liquid inertia), will be presented. It will be
restricted to the plug-flow limit ($b \gg R$), which is relevant to the case of
water inside CNTs, but could easily be generalized.
The pressure drop along the liquid inside the tube is given by the
Laplace pressure difference across the curved liquid/vapor meniscus,
minus the pressure drop at the tube entrance \cite{Sampson1891}: 
$\Delta p = 2 \Delta \gamma / R - 3 \pi \eta V / R$. If we neglect
liquid inertia, the corresponding velocity in the plug-flow 
regime is: $V = R \Delta p / (2\lambda L)$. The resulting
equation: 
\begin{equation}
2 \lambda L \od{L}{t} + 3 \pi \eta \od{L}{t} = 2 \Delta \gamma
\end{equation}
can be integrated into: 
\begin{equation}
\lambda L^2 + 3 \pi \eta L = 2 \Delta \gamma\ t .
\end{equation}
This simple quadratic equation admits the following solution: 
\begin{equation}\label{eq:fullsolution2}
\frac{L (t)}{\Lc} = \frac{1}{2} \left( \sqrt{ 1 + \frac{4t}{\tc}} - 1 \right) , 
\end{equation}
where: 
\begin{equation}\label{eq:tc2}
\tc = \frac{(3 \pi \eta)^2}{2 \lambda \Delta \gamma} 
\end{equation}
and: 
\begin{equation}\label{eq:Lc2}
\Lc = 3 \pi \eta / \lambda = 3 \pi b . 
\end{equation}
For $t \ll \tc$, the full solution \eqref{eq:fullsolution2} simplifies
to give a filling at constant velocity $V = 2\Delta \gamma / (3 \pi
\eta)$, which is compatible with the simple dimensional analysis
presented above. For $t \gg \tc$ it reduces to
Eq. \eqref{eq:washburnplug}. Interestingly, even in the plug-flow
limit, the transition length between the two regimes provides a direct
measure of the slip length (Eq. \ref{eq:Lc2}). Furthermore, for slip
lengths of a few hundredth of nanometers
\cite{Holt2006,Falk2010}, $\Lc$ would lie within the micrometer range.

\section{Conclusion}
In this article, the consequences of the giant slip of water
inside carbon nanotubes on the dynamics of capillary
filling have been investigated.
It has been shown that
in case the slip length is much larger than the tube radius (plug-flow
limit), viscosity does not play a role, and the relevant 
physical parameter controlling the flow is the liquid/solid friction
coefficient. Moreover, scaling of the various quantities as a function
of the tube radius are modified. 
It was then 
shown, using MD simulations, that in the short-time limit the
filling velocity is not limited by the liquid inertia, but rather by
viscous dissipation at the tube entrance. 
A model of the full capillary filling dynamics, taking into account
entrance effects (but neglecting liquid inertia) was finally
presented.

Generally, friction of water
inside CNT is so small that for any real flow
experiments, one can expect that the hydrodynamic resistance at the
tube input and output will play a significant role \cite{Sisan2011}. It could even 
contribute much more strongly to the total pressure drop than the
friction inside the tube. In any case, measurement of flow rate versus
pressure drop for CNTs of a given length can only give a higher limit for
the friction coefficient inside the tube, so that measurements with
different tube lengths could prove necessary.

\begin{acknowledgments}
LJ thanks A.-L. Biance, L. Bocquet, K. Falk, S. Merabia, and
O. Pierre-Louis for sharing data and/or for useful exchanges. This
work was funded by the MIKADO grant of the French Agence Nationale de la
Recherche. 
\end{acknowledgments}

%\bibliography{laurent,laurent1}

%

\end{document}